\documentclass[l10pt,letterpaper,twocolumn]{article}   
\usepackage{ol2}  
\usepackage{amsmath}
\usepackage{hyperref}
\usepackage{bm}
\usepackage{amscd}
\usepackage{amssymb}
\usepackage{graphicx}
\usepackage{epsfig}
\begin{document}
\twocolumn[ 
\title{Poincar\'e sphere representation for classical inseparable Bell-like states of the electromagnetic field}
%
%
%
\author{Annemarie Holleczek$^{*1, 2}$, Andrea Aiello$^{1,2}$, Christian Gabriel$^{1,2}$, Christoph Marquardt$^{1,2}$, Gerd Leuchs$^{1,2}$}
\address{$^1$ Max Planck Institute for the Science of Light, G\"{u}nter-Scharowsky-Str. 1/Bau 24, 91058 Erlangen, Germany}
\address{$^2$Institute for Optics, Information and Photonics, University of Erlangen-Nuremberg,\\ Staudtstr. 7/B2, 91058 Erlangen, Germany}
\address{$^*$Corresponding author: annemarie.holleczek@mpl.mpg.de}
\begin{abstract}
Classical beams of light with non-uniform polarization patterns (e.g. radially and azimuthally polarized doughnut beams) may exhibit quantum-like features as, for instance, inseparability. We establish an exact correspondence between radially and azimuthally polarized classical modes of the electromagnetic field and the  two-qubit quantum Bell states. We demonstrate the existence of a special representation for such classical modes by means of a pair of Poincar\'e spheres. Points on these spheres are described by Stokes parameters associated with such modes, and their explicit expressions are given.
\end{abstract}

\ocis{260.5430, 350.0350.}


 ] 

\maketitle 
%
%
%

Cylindrically polarized states of light with a complex polarization structure, such as radially and azimuthally polarized vector beams \cite{RadAzi1, Sheppard}, have extraordinary attributes in both the classical and quantum domain. For instance, such vector beams can be sharply focussed in the center of the optical axis and therefore, find applications in lithography, confocal microscopy and optical data storage  \cite{classical}. In the quantum regime, specially designed spatio-polarization modes can increase the coupling to single ions \cite{FourPiPAC}. Recently, it has been shown that by squeezing azimuthally polarized optical beams one can generate quantum states exhibiting \emph{hybrid} entanglement between the spatial and the polarization degrees of freedom \cite{christian}. Analogously to the polarization \cite{Bowen} and the spatial \cite{Lassen} cases, hybrid entanglement also requires the measurement of appropriate (hybrid) Stokes parameters in order to fully quantify quantum correlations.
\begin{figure}[!ht]
\centering
\includegraphics[width=8.5cm]{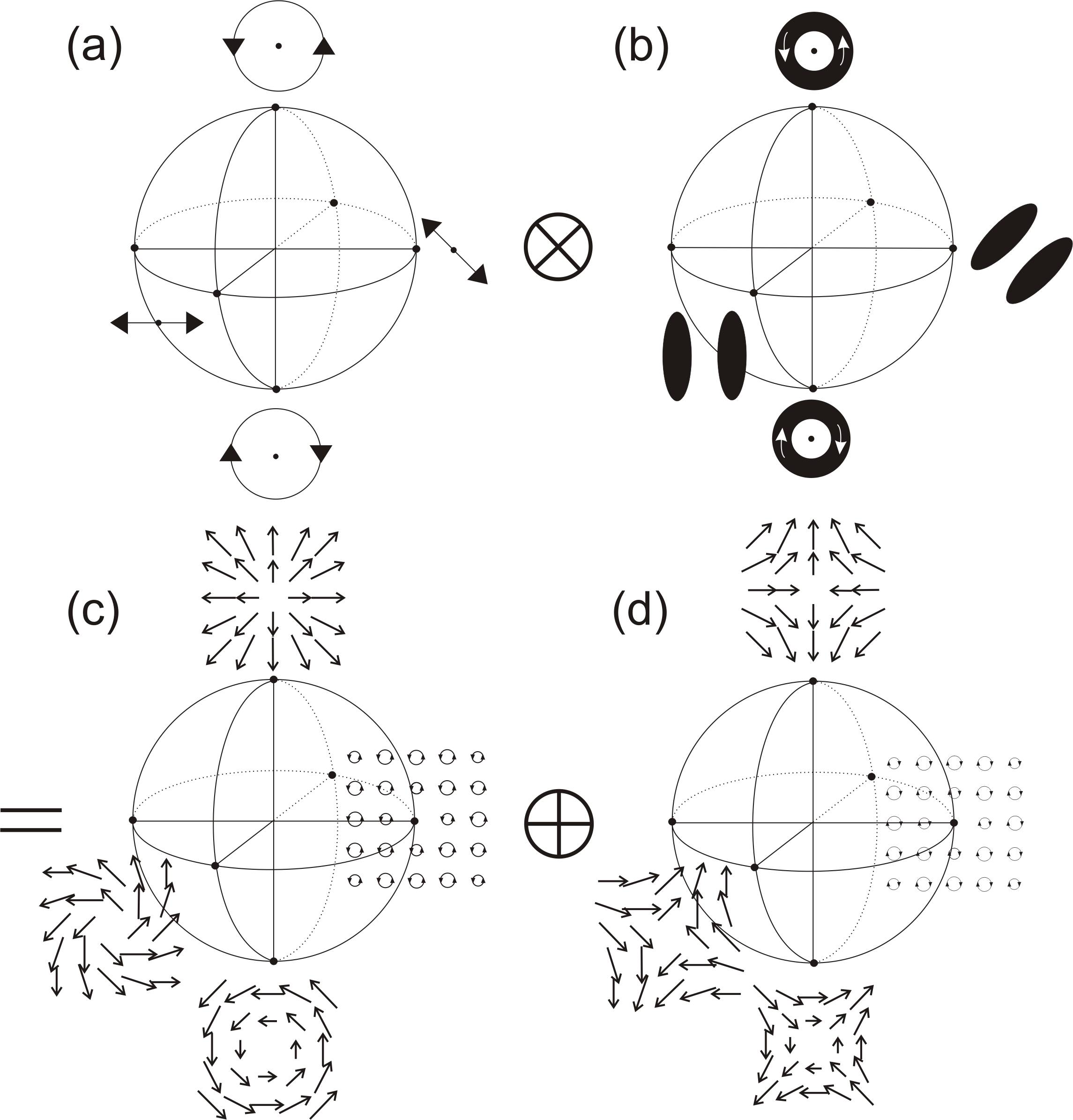}
\caption{Fundamental idea behind the new Poincar\'e sphere representation: Combining the PPS (a) and the SMPS (b) to the hybrid Poincar\'e spheres (HPSs) (c) and (d).}
\label{fig:kombination}
\end{figure}

In this Letter we aim at establishing a proper theoretical framework for the classical optics description of such Stokes parameters and to furnish their representation on a suitably defined pair of Poincar\'e spheres. This requires the introduction of four orthogonal states, each representing a cylindrically polarized light beam. Interestingly enough, these \emph{classical} states of lights are of Bell-like type, namely they have the same mathematical structure of the four \emph{quantum} Bell states well known in quantum information theory \cite{Nielsen}. The presented formalism is of utility to both the quantum and the classical communities in view of the recently growing interest in theory and applications of light beams with complex spatio-polarization patterns.

\begin{figure}
\centering
\includegraphics[width=4cm]{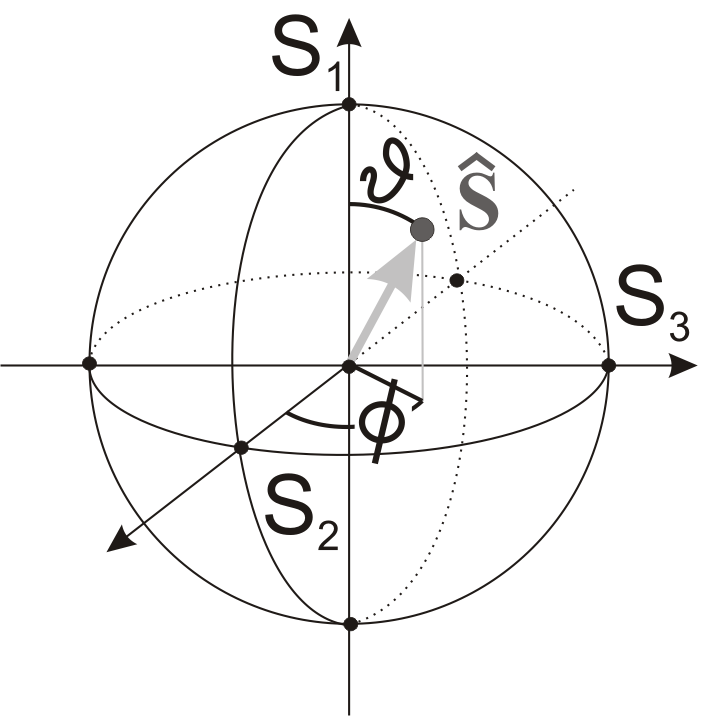}
\caption{Poincar\'e sphere representation for an arbitrary two-dimensional system. Here, the radius $r$ of the sphere is fixed to one, and spherical angles $\{\vartheta, \phi\}$ are related to the Stokes parameters via the relation: 
	$\{\cos\vartheta,\sin\vartheta \cos\phi, \sin\vartheta\sin\phi\} = \{S_1/S_0, S_2/S_0,S_3/S_0\}$, where $S_0$ denotes the total intensity of the beam.}
\label{fig:poincare}
\end{figure}

Fully polarized optical beams, similarly to quantum pure states \cite{jauch}, can be described by points on a sphere of unit radius \cite{AzzamBook}, the so-called polarization Poincar\'e sphere (PPS, Fig. \ref{fig:kombination} (a)). The North and South Poles of this sphere represent left-and right-handed polarization states that correspond to $\pm1$ \emph{spin} angular momentum eigenstates, respectively. Similarly, the spatial-mode Poincar\'e sphere (SMPS, Fig. \ref{fig:kombination} (b)) \cite{PadgettPaper} sticks to the same superposition principle but in terms of the spatial distribution of the optical beams. In this case, the North and the South Pole of the sphere represent optical beams in the Laguerre-Gaussian modes $\text{LG}_{p=0}^{\ell =\pm1}$, which correspond to $\pm1$ \emph{orbital} angular momentum eigenstates \cite{Allen1992}, respectively.
Each point on the Poincar\'e sphere, both on the polarization and the spatial-mode one, may be put in one-to-one correspondence with the three Stokes parameters $\{S_1, S_2, S_3\}$, as illustrated in Fig. \ref{fig:poincare} \cite{Stokes}.

In the following, we show that, for cylindrically polarized vector beams, it is possible to simultaneously use both the PPS and SMPS representations to obtain the two \emph{hybrid} Poincar\'e spheres (HPSs), shown in Fig. \ref{fig:kombination} (c) and (d). The North and South Pole of sphere (c) represent radially and azimuthally polarized states whose total angular momentum (orbital plus spin) is equal to zero. Analogously, sphere (d) represent counter-radial and counter-azimuthal states, still with total zero angular momentum. Each of these two pairs of states represent a novel kind of \emph{hybrid} degree of freedom (DOF) of the electromagnetic field, since such states describe neither purely polarized nor purely spatial modes of the field.

Mathematically, a combination of the PPS and the SMPS is realized by forming the four dimensional basis given by the Cartesian product of the polarization $\{ \mathbf{\hat x}, \mathbf{\hat y}\}$ and the mode $\{  \psi_{10}, \psi_{01}  \}$ bases:
\begin{equation}
\{ \mathbf{\hat x}, \mathbf{\hat y}\} \otimes \{  \psi_{10}, \psi_{01}  \}=\{  \psi_{10}\mathbf{\hat x},  \psi_{10}\mathbf{\hat y}, \psi_{01}\mathbf{\hat x}, \psi_{01}\mathbf{\hat y} \}.
\label{space}
\end{equation} 
By applying a unitary transformation to such a basis, we obtain a new set of basis vectors, namely $\{\mathbf{u}_R^+,\mathbf{u}_A^+,\mathbf{u}_R^-,\mathbf{u}_A^-\}$, where 
\begin{equation} \label{eqn:uAuR}
\mathbf{u}_R ^+ = \frac{1}{\sqrt{2}} (\mathbf{\hat x} \psi_{10} +  \mathbf{\hat y} \psi_{01} ), \,\mathbf{u}_A ^+ = \frac{1}{\sqrt{2}} (-  \mathbf{\hat x} \psi_{01} + \mathbf{\hat y} \psi_{10} ),
\end{equation}
\begin{equation}\label{eqn:uAMuRM}
\mathbf{u}_R^- = \frac{1}{\sqrt{2}} ( -\mathbf{\hat x} \psi_{10} +  \mathbf{\hat y} \psi_{01} ), \, \mathbf{u}_A^- =  \frac{1}{\sqrt{2}} (\mathbf{\hat x} \psi_{01} +  \mathbf{\hat y} \psi_{10}),
\end{equation}
with $ \psi_{nm}\equiv  \psi_{nm}(x,y,z)$, $n, m \in \mathbb \{0,1,2,\ldots\}$ being the Hermite-Gaussian solutions of the paraxial wave equations and the quantity $N = n + m$ is referred to as the order of the mode \cite{PadgettPaper}.  Here, $ \psi_{10} $ and $ \psi_{01} $ represent the two Hermite-Gaussian modes of the order $N = 1$, while $\mathbf{\hat x}$ and $\mathbf{\hat y}$ are unit vectors representing linear polarization along the $x$ and $y$ axes respectively.
The spatio-polarization patterns for $\mathbf{u}_R ^+, \,\mathbf{u}_A ^+$, $\mathbf{u}_R ^-, \,\mathbf{u}_A ^- $ are shown in Fig. 3 (a), (b), (c) and (d), respectively. 

The four basis vectors $\mathbf{u}_R ^+, \,\mathbf{u}_A ^+,\mathbf{u}_R ^-, \,\mathbf{u}_A ^- $ are clearly \emph{nonseparable}, in the sense that it is not possible to write any of these vector functions as a product of a uniform polarization vector field times a scalar function. More specifically, if one establishes the following formal equivalence between the two qubits $(A,B)$ Hilbert space and our two degrees of freedom (polarization and spatial mode) space: $|0_A\rangle\sim \mathbf{\hat x},\, |1_A\rangle \sim \mathbf{\hat y}$, $|0_B\rangle\sim \psi_{10},\, |1_B\rangle \sim \psi_{01}$, then it is not difficult to show that our basis is mathematically equivalent to the quantum Bell basis \cite{Nielsen}:
\begin{equation}
\{ \mathbf{u}_R ^+, \,\mathbf{u}_A ^+,\mathbf{u}_R ^-, \,\mathbf{u}_A ^- \} \sim \{ |\Phi ^+ \rangle,  -|\Psi^- \rangle, - |\Phi ^- \rangle,  |\Psi ^+ \rangle    \},
\end{equation} 
where $ |\Phi ^\pm \rangle = \left (|0_A,0_B\rangle \pm |1_A,1_B\rangle\right)/\sqrt{2}$ and $|\Psi ^\pm \rangle = \left (|0_A,1_B\rangle \pm |1_A,0_B\rangle\right)/\sqrt{2}$.

It is interesting to note that besides the quantum-classical analogy established above, the two sets of basis vectors $\{ \mathbf{u}^+_A,\,\mathbf{u}^+_R,\}$ and $\{\mathbf{u}^-_A,\,\mathbf{u}^-_R \}$ are connected via the following fundamental \emph{global} rotation law:
\begin{equation}
\mathbf{u}^{\pm}(\mathbf{r}, \theta \pm \alpha, z) = R(\alpha)\mathbf{u}^{\pm}(\mathbf{r}, \theta, z),
\label{rot}
\end{equation}
which states that the ``$+$'' (``$-$'') basis vectors co-rotate (counter-rotate) along with the global rotation performed by the operator $R(\alpha)$.
\begin{figure}
\includegraphics[width=8.5cm]{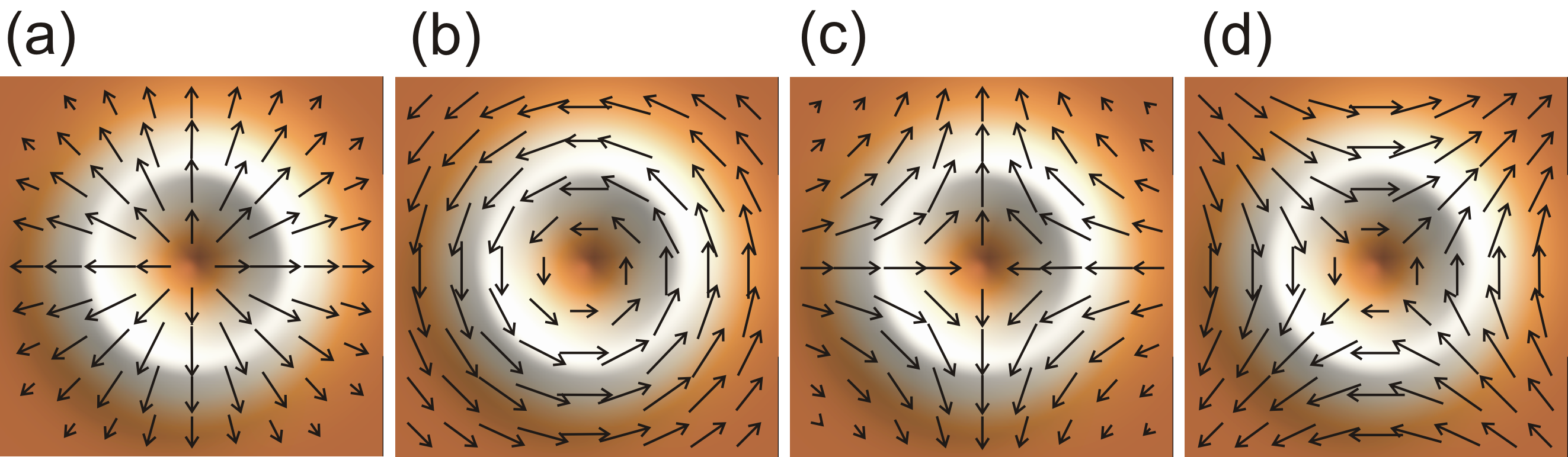}
\caption{(Color online) Complex polarization patterns of (a) $\mathbf{u}^+_R$, (b) $\mathbf{u}^+_A$, (c) $\mathbf{u}^-_R$, (d) $\mathbf{u}^-_A$, underlayed with the doughnut shaped intensity distribution.}
\label{fig:basis}
\end{figure}
We use this law as a ``superselection rule'' \cite{superselection} to split the four dimensional space $\{  \psi_{10}\mathbf{\hat x},  \psi_{10}\mathbf{\hat y}, \psi_{01}\mathbf{\hat x}, \psi_{01}\mathbf{\hat y} \} $ into the Cartesian sum of two subspaces spanned by the ``$+$'' and ``$-$'' sets of modes:
\begin{equation}
\{ \mathbf{\hat x}, \mathbf{\hat y}\} \otimes \{  \psi_{10}, \psi_{01}  \} = \{\mathbf{u}_R^+,\mathbf{u}_A^+\} \oplus \{\mathbf{u}_R^-,\mathbf{u}_A^-\}.
\end{equation}
This shows, that depending upon their behavior under global rotation, cylindrically polarized optical beams may be subdivided in two independent sets which can be represented by points on the surface of two distinct hybrid Poincar\'e spheres, as illustrated in Fig. \ref{fig:kombination}. 
It is worth to stress that a beam represented by an arbitrary superposition of either the standard basis vectors $\{  \psi_{10}\mathbf{\hat x},  \psi_{10}\mathbf{\hat y}, \psi_{01}\mathbf{\hat x}, \psi_{01}\mathbf{\hat y} \}$ or the cylindrical basis vectors $\{\mathbf{u}^+_R, \mathbf{u}^+_A, \mathbf{u}^-_R, \mathbf{u}^-_A\}$ needs four complex numbers to be described properly. This amounts to eight real numbers which can be reduced to seven due to normalization. However, we want to restrict ourselves to cylindrically polarized modes. So, in these cases, interference between ``$+$'' and ``$-$'' modes can not occur, as implied by the superselection law in Eq. (\ref{rot}). This reduces the number of real parameters necessary to describe our cylindrically polarized beams further to $3\oplus3$. This permits the introduction of our ``two-Poincar\'e Cartesian sum'' representation.

To complete our discussion about the properties of the ``$+$'' and ``$-$'' vector bases, we note that Eq. (\ref{rot}) implies that these two sets of modes do not mix under global rotation. 
From Eq. (\ref{eqn:uAuR}) it is also clear that $\mathbf{u}_R^+$ and $\mathbf{u}_A^+$ are connected by a \emph{local} counterclockwise $90^{\circ}$ rotation $D(\alpha=\pi/2)$. This local operation performed by $D(\alpha=\pi/2)$ should not be confused with the global rotation operated by $R(\alpha=\pi/2)$. While $R(\alpha)$ acts upon the polarization pattern as a whole, $D(\alpha)$ rotates each vector of the beam polarization pattern locally by the same angle $\alpha$ irrespective of its position $(x,y)$ within the pattern. A similar consideration holds for the connection between $\mathbf{u}_R^-$ and $\mathbf{u}_A^-$ if one replaces $D(\pi/2)$ with $D(-\pi/2)$. Finally, we notice that the two sets $\{\mathbf{u}_R^+,\mathbf{u}_A^+\}$ and $\{\mathbf{u}_R^-,\mathbf{u}_A^-\}$ are connected to each other by a mirror image symmetry operation that can be practically performed by a halfwaveplate \cite{DamaskBook}.

\begin{figure}
\centering
\includegraphics[width=8.5cm]{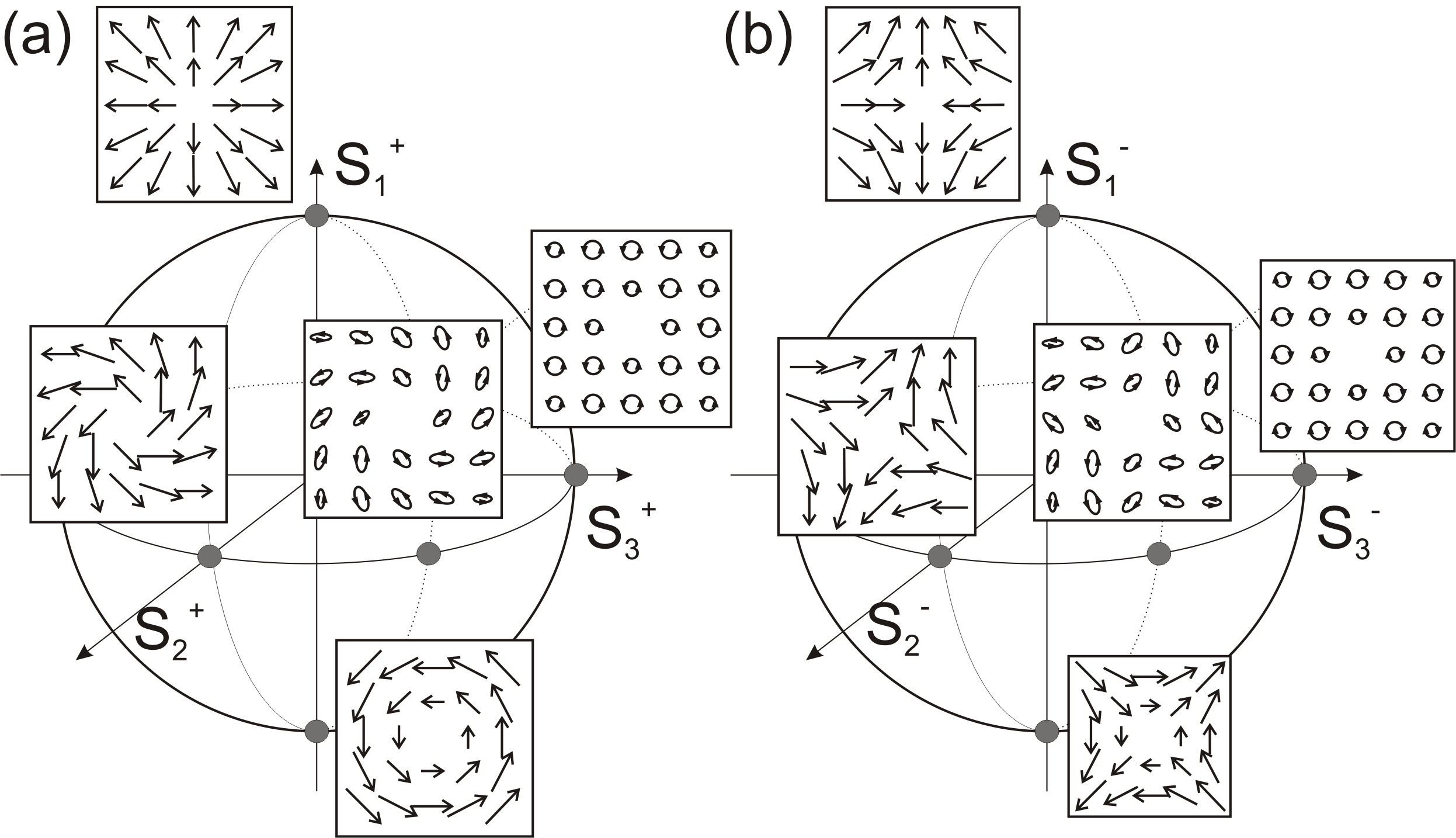}
\caption{Polarization states on the HPSs, represented on the sphere of the ``$+$'' modes (a) and the ``$-$'' modes (b).}
\label{fig:hauptpunkte}
\end{figure}

Having established a proper representation for cylindrically polarized states of light, now we explicitly describe these states in terms of \emph{hybrid} Stokes parameters. Differently from either polarization or spatial Stokes parameters, hybrid Stokes parameters convey information about both polarization and spatial-mode degrees of freedom simultaneously. They are naturally defined as 
\begin{align}
S_0 ^{\pm}= & {f_R^{\pm}}^* f_R^{\pm} + {f_A^{\pm}}^* f_A^{\pm}, \qquad \nonumber
S_1^{\pm} = {f_R^{\pm}}^* f_R^{\pm} - {f_A^{\pm}}^* f_A ^{\pm},\nonumber\\ 
S_2^{\pm} =  &{f_R^{\pm}}^* f_A^{\pm} +{ f_A^{\pm}}^* f_R ^{\pm}, \nonumber\qquad
S_3^{\pm} =-i(   {f_R^{\pm}}^* f_A^{\pm}-  {f_A^{\pm}}^* f_R^{\pm}), \nonumber
\end{align}
where the symbol $\/^*$ denotes the complex conjugation. $f_A^{\pm} = (\mathbf{u}^{\pm}_A, \mathbf{E})$ and $ f_R^{\pm} = (\mathbf{u}^{\pm}_R, \mathbf{E})$ are the field amplitudes of the electric field vector in the bases $\{ \mathbf{u}^{\pm}_A, \mathbf{u}^{\pm}_R \}$, where $\mathbf{E} = f_A^{\pm} \mathbf{u}^{\pm}_A+ f_R^{\pm} \mathbf{u}^{\pm}_R$.
These amplitudes can be expressed in terms of the spherical coordinates on the two independent hybrid Poincar\'e spheres (HPSs) as $f_A ^{\pm}= \cos{(\vartheta/2)}$ and $f_R^{\pm} =   \exp(i\phi)\sin{(\vartheta/2)}$. The first sphere of the HPSs represents the ``$+$'' modes (Fig. \ref{fig:hauptpunkte} (a)) and the second the ``$-$'' (Fig. \ref{fig:hauptpunkte} (b)) modes. It is possible to show that these Stokes parameters can be actually measured by means of conventional optical elements \cite{Holleczek}. This characteristic is particularly relevant for the possible quantum applications of our formalism, where simultaneous measurability of Stokes parameters describing spatially separated optical beams, is crucial \cite{korolkova}.

Another interesting feature of our HPSs is that every point on the meridian between the North and the South Pole ($\vartheta^{\pm}\in[ 0 , \pi] $, $\phi^{\pm} = 0$) describes a linear polarization state as illustrated in both Figs. \ref{fig:hauptpunkte} (a) and (b). Except for these points and the two ones on the $S_3$ axis ($ \vartheta = \frac{\pi}{2}, \phi = \{\frac{\pi}{2},\frac{3}{2}\pi \}$) which represent circularly polarized states, all remaining points on the spheres describe states of non-uniform elliptical polarization.

To summarize, a hybrid representation for cylindrically polarized vector beams has been presented in analogy to the well-known Poincar\'e sphere representations for the polarization and the orbital angular momentum degrees of freedom. Explicit expressions for the Stokes parameters representing these states of the electromagnetic field are given. Moreover, we were able to show that it exists an exact formal analogy between such cylindrically polarized vector beams and maximally entangled two-qubit states (Bell states). The latter result suggests possible intriguing applications of our formalism to quantum states of light \cite{christian}.

The authors thank Peter Banzer for fruitful discussions.



\begin{thebibliography}{10}

\bibitem{RadAzi1}
R. Martinez-Herrero, P. M. Mej\'ias, G. Piquero
\newblock {\em Characterization  of Partially Polarized Light Fields}.
\newblock Springer (2009).

\bibitem{Sheppard}
C. J. R. Sheppard
\newblock {\em Polarization of almost-plane waves}.
\newblock J. Opt. Soc. Am. A $\mathbf{17}$, 335 (2000).

\bibitem{classical}
R. Dorn, S. Quabis and G. Leuchs
\newblock {\em Sharper Focus for a Radially Polarized Light Beam}.
\newblock Phys. Rev. Lett. $\mathbf{91}$, 233901 (2003).

\bibitem{FourPiPAC}
M. Sondermann, R. Maiwald, H. Konermann, N. Lindlein, U.  Peschel, G. Leuchs
\newblock {\em Design for a mode converter for efficient light-atom coupling in free space}.
\newblock Appl. Phys. B $\mathbf{89}$, 489 (2007).

\bibitem{christian}
C. Gabriel et al.
\newblock {\em Hybrid entanglement in continuous variable systems}.
\newblock arXiv:1007.13221 (2010).

\bibitem{Bowen}
W. P. Bowen, N. Treps, R. Schnabel, P. K. Lam
\newblock {\em Experimental Demonstration of Continuous Variable Polarization Entanglement}.
\newblock Phys. Rev. Lett. $\mathbf{89}$, 253601 (2002).

\bibitem{Lassen}
M. Lassen, G. Leuchs, U. L. Andersen
\newblock {\em Continuous Variable Entanglement and Squeezing of Orbital Angular Momentum States}.
\newblock Phys. Rev. Lett. $\mathbf{102}$, 163602 (2009).

\bibitem{Nielsen}
M. A. Nielsen, I. L. Chuang
\newblock {\em Quantum computation and quantum information}.
\newblock Cambridge University Press (2000).

\bibitem{jauch}
J. M. Jauch, F. Rohrlich
\newblock {\em The Theory of Photons and Electrons: The relativistic quantum Field Theory of Charged Particles with Spin one-half}.
\newblock Addison-Wesley Publishing Company (1955).

\bibitem{AzzamBook}
R. M. A. Azzam and N. M. Bashra
\newblock {\em Ellipsometry and Polarized Light}.
\newblock North Holland Personal Library (1998).

\bibitem{PadgettPaper}
M. J. Padgett, J. Courtial
\newblock {\em Poincar\'e-sphere equivalent for light beams containing orbital angular momentum}.
\newblock Opt. Lett. $\mathbf{24}$, 430 (1999).

\bibitem{Allen1992}
L. Allen, M. W. Beijersbergen, R. J. C. Spreeuw, J. P. Woerdman 
\newblock {\em Orbital angular momentum of light and the transformation of Laguerre-Gaussian laser modes}.
\newblock Phys. Rev. A $\mathbf{45}$, 8185 (1992).

\bibitem{Stokes}
G. G. Stokes
\newblock{\em On the composition and resolution of streams of polarized light from different sources}
\newblock Trans. Cambridge Philos. Soc. $\mathbf{9}$, 399 (1852).

\bibitem{superselection}
A. Peres
\newblock {\em Quantum Theory: Concepts and methods}.
\newblock Kluwer Academic, Boston (1995).

\bibitem{DamaskBook}
Jay~N. Damask.
\newblock {\em Polarization Optics in Telecommunications}.
\newblock Springer, 1 edition (2004).

\bibitem{Holleczek}
A. Holleczek {\it{et al.}}, in preparation.

\bibitem{korolkova}
N. Korolkova, G. Leuchs, R. Loudon, T. C. Ralph, C. Silberhorn
\newblock {\em Polarization squeezing and continuous-variable polarization entanglement}.
\newblock Phys. Rev. A $\mathbf{65}$, 052306 (2002).

\end{thebibliography}
\end{document}